\documentclass[12pt,twoside]{article}
\usepackage{fleqn,espcrc1}

\usepackage{graphicx}
\usepackage[figuresright]{rotating}

\newcommand{\AmS}{{\protect\the\textfont2
  A\kern-.1667em\lower.5ex\hbox{M}\kern-.125emS}}

\title{Percolation and Deconfinement in SU(2) Gauge Theory}

\author{S. Fortunato\address{Fakult{\"a}t f{\"u}r Physik, 
    Universit{\"a}t Bielefeld,\\
    D-33501 Bielefeld, Germany}%
        \thanks{The work has been supported by the TMR network ERBFMRX-CT-970122
  and the DFG under grant Ka 1198/4-1.}
        and 
        H. Satz$^{\mathrm a}$
        }
       
\begin{document}

% typeset front matter
\maketitle

\begin{abstract}
The deconfinement transition in SU(2) gauge theory and the 
magnetization transition in the Ising model belong to the 
same universality class. The critical behaviour of the Ising model
can be characterized either as spontaneous breaking of the $Z_2$ symmetry 
of spin states or as percolation of appropriately defined spin clusters. 
We show that deconfinement in SU(2) gauge theory can be specified as 
percolation of Polyakov loop clusters with Fortuin-Kasteleyn bond weights,
leading to the same critical exponents as the
conventional order-disorder description based on the 
Polykov loop expectation value.
\end{abstract}

\section{INTRODUCTION}

The study of critical phenomena has always been one of the 
most challenging and fascinating topics in physics.
One can give many examples of systems which 
undergo phase transitions, from familiar 
cases like the boiling of
water in a kettle or the 
paramagnetic-ferromagnetic transition of 
iron at the Curie temperature, to 
the more complicated case of 
the transition from hadronic 
matter to quark-gluon plasma which is likely
to be obtained by high-energy heavy-ion experiments 
in the coming years. 
Particularly interesting are
the second-order phase transitions, characterized
by a continuous variation 
of the order parameter and 
a divergent correlation length at
the threshold. Already in the 70's 
one tried to use percolation
theory in order to give 
a geometrical description of 
the critical behaviour of dynamical systems 
undergoing second-order phase transitions \cite{krumba,Sykes,Co1}.
Percolation has in fact several
features in common with such systems: power law behaviour 
of the variables at criticality, scaling relations, universality.
  
The critical behaviour of the Ising model can indeed be
reformulated in terms of percolation theory: magnetization sets in
when suitably defined clusters of parallel spins reach the dimensions
of the system \cite{Co80}. 
In particular, the critical exponents for percolation
then become equal to the Ising exponents.

The paramagnetic-ferromagnetic transition of the 
Ising model is strictly related to the confinement-deconfinement
transition in finite
temperature $SU(2)$ gauge theory. This was conjectured on the 
basis of effective theories
\cite{Janos,S&Y} and confirmed by lattice studies \cite{Engels}.
The order parameter for $SU(2)$ is the lattice average of the Polyakov loop,
and it behaves like the magnetization in the Ising model; in particular,
the critical exponents are the same.

These analogies are the basis of this work. We have tried to see whether
it is possible to find a description of critical behaviour in terms of
percolation also for the deconfinement transition in $SU(2)$ gauge theory.
We show that in a lattice 
regularization which corresponds
to the strong coupling limit, both in two and in three space dimensions,
the percolation of Polyakov loop clusters (taken to be suitably defined 
areas of Polyakov loops $L$ of the same
sign) leads to the correct deconfinement temperature and to the correct
critical exponents for the deconfinement. 
We have achieved this result by means of computer simulations
of finite temperature $SU(2)$ gauge theory and using
standard finite size scaling tecniques to extrapolate 
the results to the infinite volume limit.

\section{PERCOLATION THEORY AND THE ISING MODEL}

    The percolation problem \cite{Stauff,Grimm} is easy to formulate: 
    just place randomly pawns on a chessboard.
    Regions of adjacent pawns form 
    {\it clusters}.  Percolation theory deals with
    the properties of these clusters when the chessboard is infinitely
    large. If one of the clusters spans the chessboard from 
    one side to the opposite one, we say that the cluster 
%\newpage
\noindent{\it percolates}. 

%     \begin{figure}[htb]
%       \vspace{5pt}
%       \epsfig{file=config_rand.eps,width=55mm}
%       \caption{2-dimensional site percolation on a square lattice.}
%       \label{fig:ranperc}
%     \end{figure}

    Quantitatively, one counts how many pawns belong 
    to each cluster and calculates two quantities:\\
      {$\bullet$}  The {\it average cluster
        size } S, defined as:
      
      {\begin{center}
        \begin{equation}
          S={\sum_{s} \Bigg({{n_{s}s^2}\over{\sum_{s}{n_{s}s}}}\Bigg)}~.
        \end{equation}
        \end{center}}
      
      Here $n_{s}$ is the number of clusters of size $s$
      and the sums exclude the percolating cluster; this number 
      indicates how big on average the clusters are which do not
      percolate.\\
      {$\bullet$} The {\it percolation strength } P, 
      defined as:
      {
        \begin{equation}{
            P=\frac{\mbox{\it size of the percolating cluster}}
            {\mbox{\it no. of lattice sites}}}~.
        \end{equation} }
      \noindent 

      By varying the density of our pawns, a kind of phase transition occurs. 
      We pass from a phase of
      non-percolation to a phase in which one of the clusters percolates.
      The percolation strength P is the {\it order parameter}  of this
      transition: it is zero in the non-percolation phase and is different from
      zero in the percolation phase. 
      
      If we want to study percolation in dynamical systems,
      for example the Ising model, first of all
      we must define the rule
      to build up the clusters (for instance
      we can join together all nearest-neighbour 
      spins of the same sign). Then one has to find out at which
      temperature we have a spanning cluster. 
      Let us call this temperature $T_p$. It turns out that 
      for $T{\approx}T_p$ the 
      percolation variables P and S behave in the following way:
      
      \begin{equation}
      \,\,\,\,\,\,\,\,\,\,\,\,\,\,\,\,\,\,\,\,\,\,\,\,\,\,\,\,\,\,\,\,\,\,\,\,\,\,\,\,\,\,\,\,\,\,\,\,\,\, \,\,\,\,\,\,\,\,\,
       P~{\sim}~(T_p-T)^{\beta_p}, 
\,\,\,\,\,\,\,\,\,\,\,\,\,\,\,\,\,\,\,\,\,\,\,\,\,\,\,\, {S}~
        {\sim}~|T-T_p|^{-\gamma_p}
      \end{equation}
        
      A. Coniglio and W. Klein \cite{Co80} demonstrated that 
      for some special definition of cluster, the onset of percolation
      coincides with the one of magnetization; besides,  
       $\beta_p$ and $\gamma_p$
      coincide respectively with 
      the magnetization exponent $\beta$
      and the susceptibility exponent $\gamma$.
      Such cluster definition had already
      been used by Fortuin and Kasteleyn to show that
      the partition function of the Ising 
      model can be rewritten in purely geometrical terms
      as a sum over clusters configurations \cite{FK}.
      According to the Fortuin-Kasteleyn prescription,
      two nearest-neighbouring spins of the same
      sign belong
      to the same cluster with a probability $p=1-exp(-2{\beta})$
      ($\beta=J/kT$, J is the Ising coupling, T the temperature).
      The result of Coniglio and Klein is valid
      in any space dimension $d$ ($d{\geq}2$) and it
      is independent of the lattice geometry 
      (square, triangular, honeycomb, etc.)
      as long as it is 
      homogeneous.

\section{POLYAKOV LOOP PERCOLATION IN SU(2) GAUGE THEORY}

Finite temperature 
$SU(N)$ gauge theories describe the 
interactions of systems of mutually interacting gluons 
in thermal equilibrium ($N$
is the number of colours). Systems containing quarks together
with gluons are certainly more appealing; 
such a scenario will be explored experimentally
up to very high temperatures
by means of high-energy heavy ion collisions.
Nevertheless, $SU(N)$ gauge theories 
are of theoretical interest; already at this
level one has a transition between 
confinement to deconfinement, going from a phase
of bound state of gluons (glueballs) to a phase
of free gluons. 
The order parameter
of the confinement-deconfinement phase transition of 
$SU(N)$ gauge theories is the lattice average of the Polyakov
loop $<L>$. It is related to the 
effective potential $V(r)$ of a static (mass $\rightarrow\infty$) 
quark-antiquark pair put in the
gluons' medium at temperature 
$T$ at a distance $r$ from each other, when $r$ is very
big:
 
\begin{equation}
  |<L>|^2~{\sim}~\lim_{r\to\infty}~{e^{-{V(r)\over T}}},
\end{equation}

The Polyakov loop is zero in the 
confined phase ($V(r)\rightarrow\infty$) and 
different from zero in the deconfined phase ($V(r)$ finite). 
Some time ago it was conjectured by B. Svetitsky and L. G. Yaffe
that the critical behaviour of 
$SU(N)$ gauge theories has a strong relationship 
with the one of simple $Z(N)$ spin models, with which
they share a common global symmetry \cite{S&Y}. 
In particular, in case of second order phase transitions,
both models would belong to the same universality class,
that is they would have the same set of critical exponents.
One simple test of the Svetitsky-Yaffe conjecture is
provided by the $SU(2)$ gauge theory: numerical simulations
showed that its critical exponents indeed coincide with the ones of  
the Ising model \cite{Engels}. 

In some respect the lattice configurations of 
$SU(2)$ are similar to the ones of the Ising model.
Instead of having a spin variable at each
lattice site we have the value of the Polyakov loop at that
site. 
Our aim is to see whether it 
is possible to build clusters of Polyakov loops
such that their percolation indices (threshold, exponents)
coincide with the thermal ones.
In order to do that, we have to face two difficulties:

i) the Polyakov loop is not a two-valued 
variable like the spin in the Ising model
but a continuous one;
its values vary in a range that, with the normalization
convention we use, is $[-1,1]$; 

ii) the $SU(2)$ Lagrangian is not directly a function of the 
Polyakov loop, therefore it is not 
possible to extract from it the expression of the
bond probability that we need to build the clusters like 
in the Ising model. 

In a recent work \cite{InfIs} it was proved that the point
i) is not a problem. If we take an Ising model with
continuous instead of two-valued spins, the definition 
of Fortuin-Kasteleyn clusters can be generalized by
taking as a bond probability between two 
positive (or negative) nearest-neighbour spins $s_i$
and $s_j$ the following expression:

\begin{equation}
p(i,j)=1-exp(-2{\beta}{s_i}{s_j})
\end{equation}

The problem ii), however, is  hard to overcome. In particular, it seems
that there is no way to express the lattice Lagrangian 
of $SU(2)$ in terms of Polyakov loops for all lattice
regularizations. Therefore we were forced to 
investigate $SU(2)$ in a special lattice regularization,
which corresponds to the so called strong coupling limit.
In this case it was shown \cite{Ka} 
that the partition function of $SU(2)$
can be written in a form which, apart 
from a factor which depends on the group
measure, is the partition function 
of the continuous Ising model studied in 
\cite{InfIs}. The coupling 
$\kappa$ of the spin model and the 
coupling $\beta$ of $SU(2)$ 
are related to each other by this relation:

\begin{equation}
\kappa \simeq ({\beta}/2)^{N_t} 
\end{equation}

\noindent($N_t$ fixes the temporal lattice regularization). 
If $SU(2)$ is approximately a continuous Ising model,
we can try to use the general definition of 
clusters of \cite{InfIs}. Therefore
we have defined our clusters as regions of like-sign
Polyakov loops connected by bonds distributed
according to the bond weight
\begin{equation}
p(i,j) = 1 - \exp(-2\kappa L_i L_j), 
\end{equation}
\noindent($L_i$ and $L_j$ are the Polyakov loop values at the sites $i$ and $j$).
For ${N_t}=2$ 
the approximation of  \cite{Ka}
is good and we 
chose to investigate numerically this special case in two
and three space dimensions. 

\section{RESULTS}

We began our percolation studies performing some test runs 
for different lattice sizes to check the behaviour of our 
percolation variables around criticality.
Figure 1 shows the average cluster size $S$ for 
$2+1 \,\,SU(2)$ in correspondence of different lattice sizes.
To get the critical point of the percolation transition we used 
the method suggested in \cite{Bin}.
For a given lattice size and 
a value of $\beta$  we counted how many times 
we found a percolating cluster. This number is successively divided
by the total number of configurations at that $\beta$.
We call this quantity percolation probability.
This variable is directly a scaling function,
analogue to the Binder cumulant 
in continuous thermal phase transitions.
Figure 2 shows the percolation probability for 
the 3-dimensional case as a function of $\beta$
for $24^3\times 2$,  
$30^3\times 2$ and $40^3\times 2$.

\begin{figure}[htb]
\begin{minipage}[t]{78mm}
\flushleft\includegraphics[width=79mm]{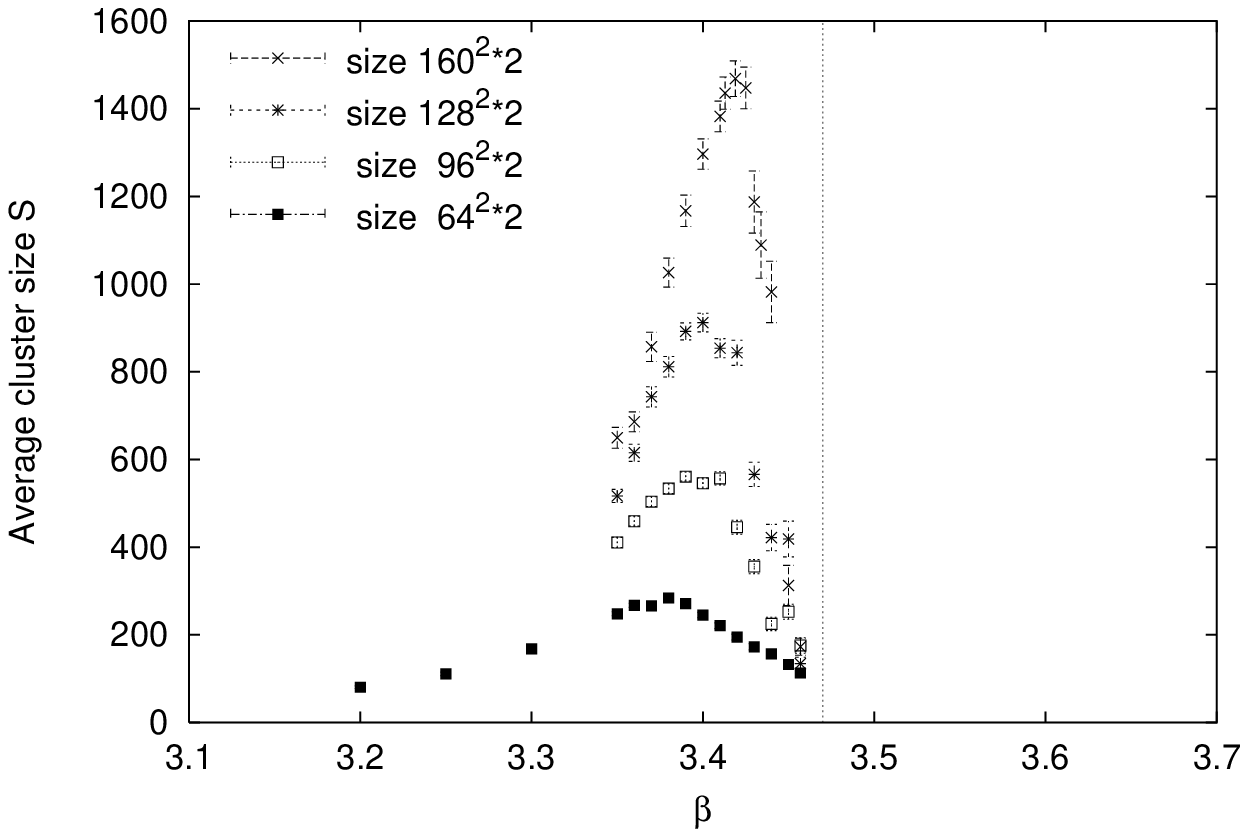}
\vskip -0.9cm
\caption{Average cluster size as a function of $\beta$ for 
$2+1 \,\,SU(2)$ and different lattice sizes. The dashed line indicates
the position of the physical threshold.}
\label{fig:largenenough}
\end{minipage}
\hspace{\fill}
\begin{minipage}[t]{78mm}
\flushright\includegraphics[width=79mm]{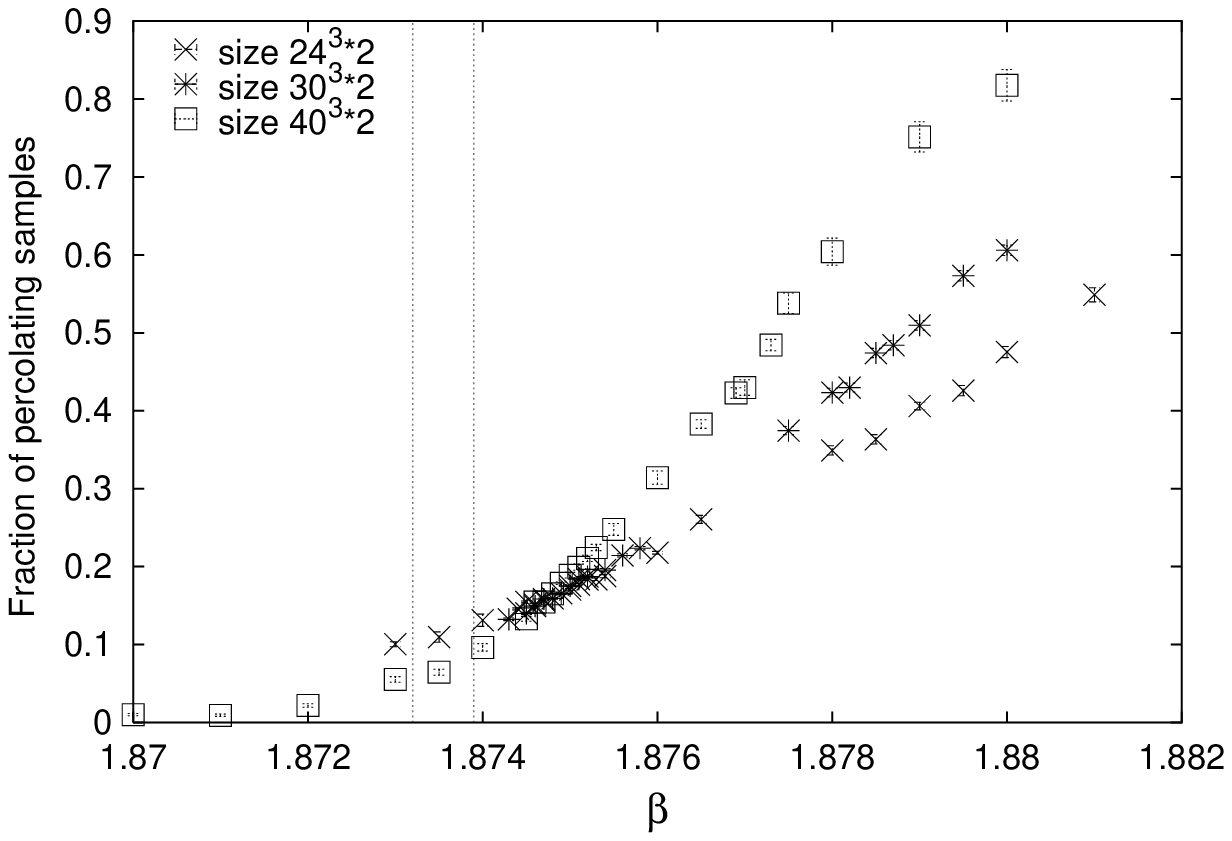}
\vskip -0.9cm
\caption{Percolation probability for $3+1 \,\, SU(2)$ as a function of $\beta$ for three 
lattice sizes.}
\label{fig:toosmall}
\end{minipage}
\vskip 0.4cm
\begin{minipage}[t]{78mm}
\flushleft\includegraphics[width=79mm]{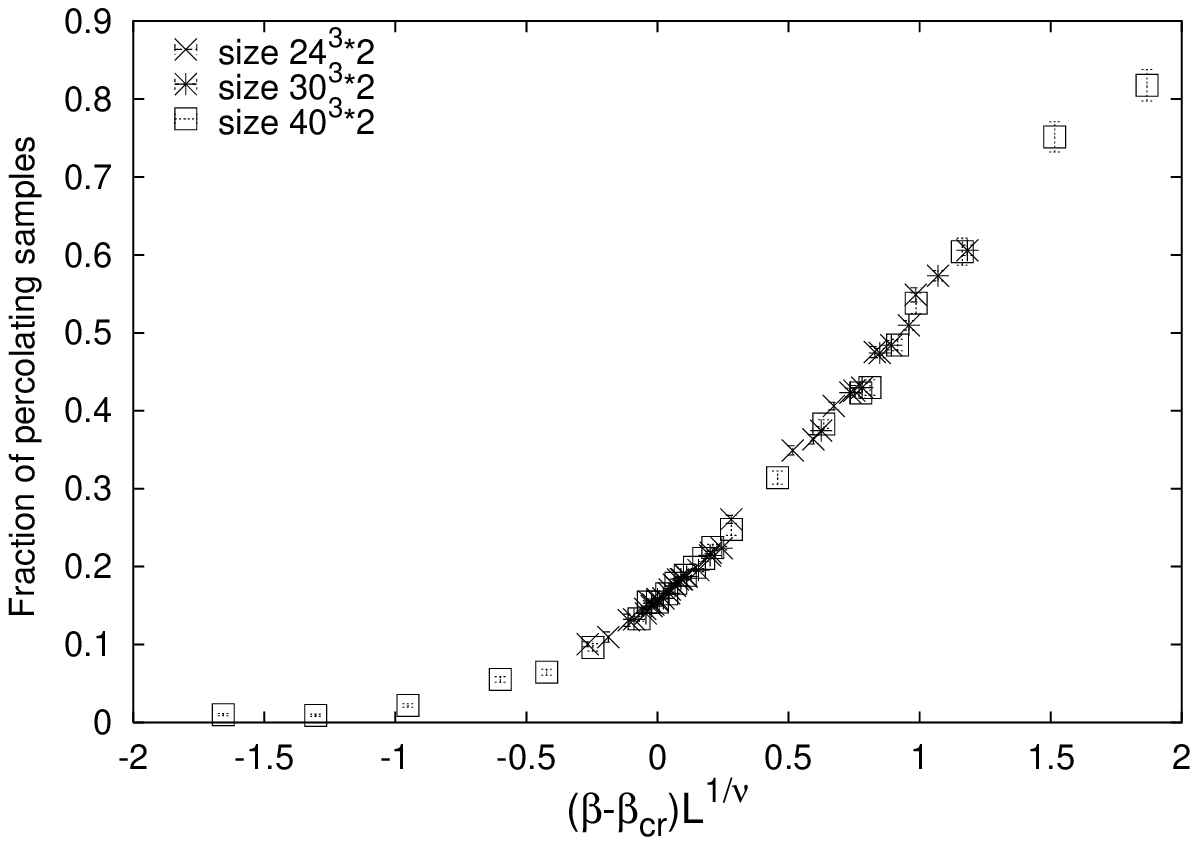}
\vskip -0.9cm
\caption{Rescaled percolation probability for $3+1 \,\, SU(2)$ using $\beta_{cr}=1.8747$
and the 3-dimensional Ising exponent
  $\nu=0.629$.}
\label{fig:largenenough}
\end{minipage}
\hspace{\fill}
\begin{minipage}[t]{78mm}
\flushright\includegraphics[width=79mm]{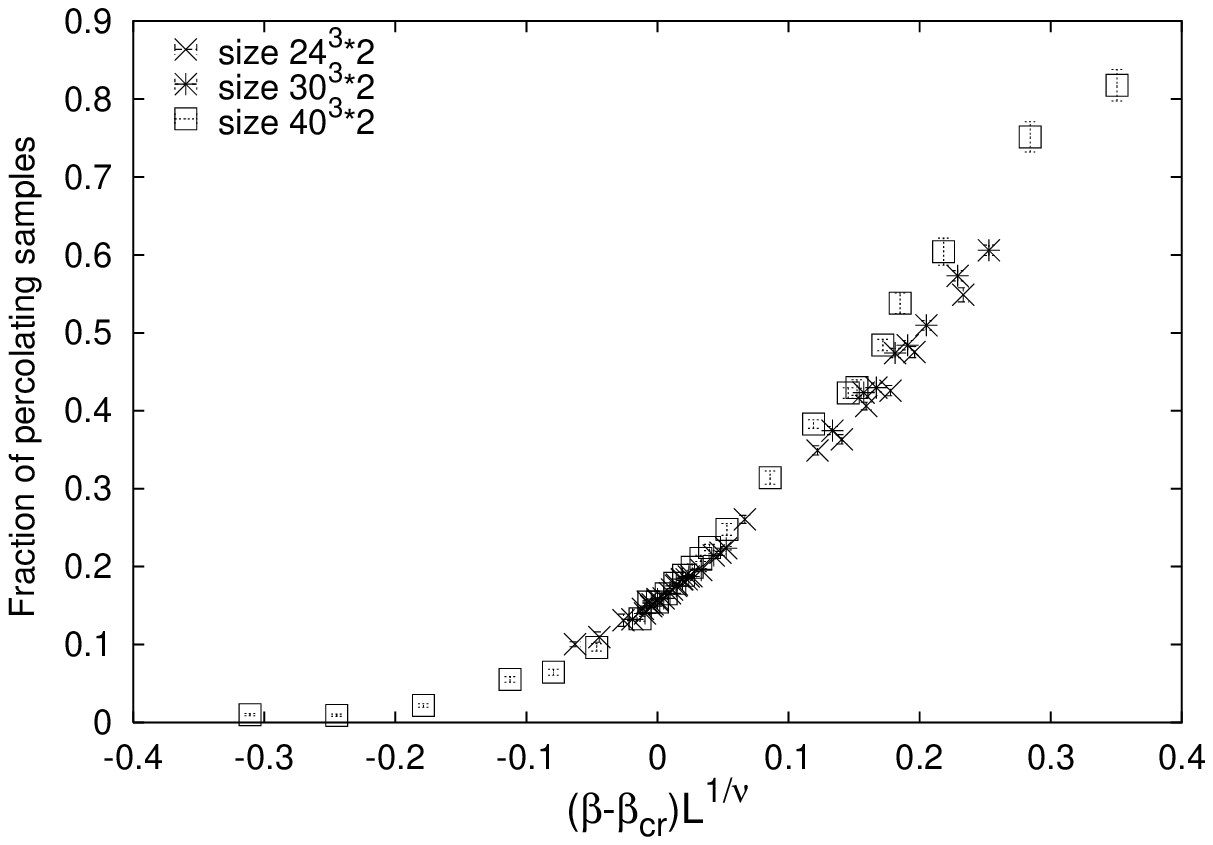}
\vskip -0.9cm
\caption{Rescaled percolation probability for $3+1 \,\, SU(2)$ using $\beta_{cr}=1.8747$
and the 3-dimensional random percolation
 exponent $\nu=0.88$.}
\label{fig:toosmall}
\end{minipage}

\end{figure}
The lines cross at the same point within the errors and 
that restricts further our $\beta$ range  
for the critical threshold.
Besides, since the percolation probability is a scaling function, we could
already get clear indications about the class of critical exponents of
our clusters.
In fact, if one knows the critical point and the exponent $\nu$,
a rescaling of the percolation probability as a function
of $(\beta-\beta_{cr})L^{1/\nu}$ should give us the same function
for each lattice size. 
Fig. 3 and 4 show the rescaled percolation probability 
for 3-dimensional $SU(2)$ using $\beta_{cr}=1.8747$
and two different values of the exponent $\nu$, respectively the
Ising value and the random percolation one.
The figures show clearly a remarkable scaling for the Ising 
exponent and no scaling for the random percolation exponent.

\vskip -0.8cm
To evaluate the exponents $\beta$ and $\gamma$ we performed high-statistics
simulations in the range where the percolation probability curves cross
each other. The number of measurements we took for each 
value of the coupling varies from 50000 to 100000.
We used the $\chi^2$ method \cite{Engels} to determine the 
values of the exponents. The final results are reported in Table 1 and 2.
The agreement both in two and in 
three dimensions is good.
\vskip 1cm
\begin{table}[htb]
\caption{Comparison of the thermal and the percolation results for 2+1 SU(2).}
\label{table:II}
\newcommand{\m}{\hphantom{$-$}}
\newcommand{\cc}[1]{\multicolumn{1}{c}{#1}}
\renewcommand{\tabcolsep}{1pc} % enlarge column spacing
\renewcommand{\arraystretch}{1.2} % enlarge line spacing
\begin{tabular}{@{}lllll}
\hline
$$            & \cc{Critical point} & \cc{${\beta}/{\nu}$} & \cc{${\gamma}/{\nu}$} & \cc{$\nu$} \\
\hline
L Percolation                 & 3.443(1)
&0.128(5) 
&1.752(8)  & 0.98(5) \\
Spont. Symm. Breaking     & 3.464(14)
&1/8  
& 7/4& 1 \\
\hline
\end{tabular}\\[2pt]
\end{table}
\vskip -1cm
\begin{table}[htb]
\caption{Comparison of the thermal and the percolation results for 3+1 SU(2).}
\label{table:I}
\newcommand{\m}{\hphantom{$-$}}
\newcommand{\cc}[1]{\multicolumn{1}{c}{#1}}
\renewcommand{\tabcolsep}{1pc} % enlarge column spacing
\renewcommand{\arraystretch}{1.2} % enlarge line spacing
\begin{tabular}{@{}lllll}
\hline
$$            & \cc{Critical point} & \cc{${\beta}/{\nu}$} & \cc{${\gamma}/{\nu}$} & \cc{$\nu$} \\
\hline
L Percolation                 & 1.8747(2)
&0.528(15) 
&1.985(13)  & 0.632(11) \\
Spont. Symm. Breaking& 1.8735(4)
&0.523(12)  
& 1.953(18)& 0.630(14) \\
Ising Model \cite{ferr} &   & 0.518(7)  &
1.970(11)  
& 0.6289(8)  \\
\hline
\end{tabular}\\[2pt]
%%The Ising values are taken from \cite{ferr}.
\end{table}

\section{CONCLUSIONS}

We have shown that the confinement-deconfinement phase transition
in finite temperature $SU(2)$ pure gauge theory
can be described as percolation of some suitably defined
clusters of Polyakov loops of the same sign. Our result is valid 
only in the strong coupling limit, and in order to make it general 
a different approach seems to be necessary. The use of effective 
theories for $SU(2)$ may help to solve the problem \cite{petr}.

\end{document}